\documentclass[twocolumn]{article}

\usepackage{scrextend}
\usepackage{lineno,hyperref}
\usepackage{threeparttable}
\usepackage{subcaption}
\usepackage{graphicx}
\usepackage{adjustbox}
\usepackage{tablefootnote}
\usepackage[affil-it]{authblk}

\begin{document}

\title{\vspace{-2.0cm}Numerical constraints on the size of generation ships from total energy expenditure on board, annual food production and space farming techniques}

\author{Fr\'ed\'eric Marin\textsuperscript{1}, Camille Beluffi\textsuperscript{2}, Rhys Taylor\textsuperscript{3} \& Lo\"{\i}c Grau\textsuperscript{4}\\
{\small 1 Universit\'e de Strasbourg, CNRS, Observatoire astronomique de Strasbourg, UMR 7550, F-67000 Strasbourg, France\\
2 CASC4DE, Le Lodge, 20, Avenue du Neuhof, 67100 Strasbourg, France\\
3 Astronomical Institute of the Czech Academy of Sciences, Bocni II 1401/1a, 141 00 Praha 4, Czech Republic \\
4 Morphosense, 18 all\'ee du Lac Saint Andr\'e, 73370 Le Bourget du Lac, France}}
              
\date{Dated: \today}

\twocolumn[
  \begin{@twocolumnfalse}
    \maketitle
    \begin{abstract}
    In the first papers of our series on interstellar generation ships we have demonstrated that the numerical
    code HERITAGE is able to calculate the success rate of multi-generational space missions. 
    Thanks to the social and breeding constraints we examined, a multi-generational crew can safely reach an 
    exoplanet after centuries of deep space travel without risks of consanguinity or genetic disorders. 
    We now turn to addressing an equally important question : how to feed the crew? Dried food stocks are not a viable
    option due to the deterioration of vitamins with time and the tremendous quantities that would be required for long-term
    storage. The best option relies on farming aboard the spaceship. Using an updated version of HERITAGE that now accounts 
    for age-dependent biological characteristics such as height and weight, and features related to the varying
    number of colonists, such as infertility, pregnancy and miscarriage rates, we can estimate the annual 
    caloric requirements aboard using the Harris-Benedict principle. By comparing those numbers with conventional 
    and modern farming techniques we are able to predict the size of artificial land to be allocated in the vessel 
    for agricultural purposes. We find that, for an heterogeneous crew of 500 people living on an omnivorous, 
    balanced diet, 0.45~km$^2$ of artificial land would suffice in order to grow all the necessary food using a 
    combination of aeroponics (for fruits, vegetables, starch, sugar, and oil) and conventional farming (for meat, 
    fish, dairy, and honey). 
    \end{abstract}    
    
    {\small {\bf Keywords:} Long-duration mission -- Multi-generational space voyage -- Space colonization -- Space settlement -- Space farming -- Space genetics}
    \vspace{3\baselineskip}
  \end{@twocolumnfalse}
]

\section{Introduction}
\label{Introduction}
The first controlled, successful airship flights were made at the end of the nineteenth century. A century later, 
humans are able to launch unmanned interplanetary probes and crewed orbital space stations in low Earth orbit. The race towards 
the stars is accelerating and technologies are evolving to allow humanity to reach for neighbouring planets \cite{Dostal2003,Perchonok2011,Rapp2015}. 
To reach more distant planets, i.e. planets orbiting around another star than the Sun, means travelling much larger distances in
deep space. To achieve such interstellar journeys in the scope of colonization, large spaceships will be necessary to transport 
human settlers. The question of the methods for ensuring a healthy population to reach its destination has been extensively covered
in the literature, and it was concluded that the only feasible scenario relies on multi-generational spaceships \cite{Moore2003,Smith2014,Marin2017}.
The initial population would grow old and die, leaving their descendants to continue travelling.

The numerical tool HERITAGE was created in 2017 to investigate the mathematical, biological, demographical and statistical 
feasibility of such an undertaking. HERITAGE is a code based on the Monte Carlo method, a mathematical approach that uses 
random draws to perform calculations, allowing tests of all possible outcomes of a given scenario by repeat iterations of the code.
This makes it possible to estimate the success rate and the associated uncertainties on the results for any 
kind of complex situation, such as the evolution of a space crew through several generations. Each iteration ​starts with 
a crew with given biological and demographical input data, allowed to vary with time, and the code simulates the billions 
of interactions that can occur between breeding partners. This can thus determine if a ship with such initial crew can 
survive without inbreeding or genetic decay for several centuries \cite{Marin2017}. We highlighted in \cite{Marin2018} 
that the most important parameters to monitor during the space travel are the inbreeding coefficient and the total 
population in the vessel. As a consequence, to maintain a genetically healthy community, the multi-generational crew must 
follow $adaptive$ social engineering principles, which means that each year these principles should be revised in order 
to ensure the success of the mission. To achieve such goal, we found that an initial ship with no less than 98 settlers 
is needed. Lower initial populations would lead to decreasing chances of mission success. We also found that it was 
mandatory to increase the allowed procreation window for women, with respect to \cite{Moore2003}, in order to keep a 
stable population level that can recover for catastrophic events that might occur during the interstellar trip.

We now turn our attention to the survivability of the population in terms of resources. Logically, our 
next step consists on the investigation of the food requirements to feed the crew. Astronauts on the International Space 
Station (ISS) require approximately 1.8 kilograms of food and packaging per day \cite{Cooper2011}. ​So if we were to feed
the crew of an interstellar mission entirely from stored food, the mass required reaches millions of tonnes. Additionally, 
the amount of vitamins contained in the food significantly decreases with time, independently of the storage processes 
\cite{Weits1970,Kalt2005,Osunde2009}. The prospect of storing the food supply required for the whole trip is therefore not viable.
In contrast, space ​agriculture, which produces fresh food, recycles nutrients and faeces, generates oxygen, 
and continuously purifies the air is by far the best option to feed a large-scale population and avoid vitamin deficiency.
In principle a space farm can transform the spaceship into a complete, closed ecological system. Studies of such systems 
already exists, ​like the Lada experiment on the ISS, running since 2002 \cite{Bingham2002,Zabel2014}. It uses a 
greenhouse-like chamber to grow plants to investigate the safety of space-grown crops, the micro-organisms they might 
have to deal with, and how to optimize crop productivity. Meat production has not been considered yet​, but the recent 
developments of artificial meat grown from cultured cells in laboratories raises the possibility that astronauts 
could avoid a purely vegan diet without the enormous associated support network necessary for animal farming 
\cite{Woll2018}.

We therefore begin our investigation into the required resources by examining the necessities of food production. How big should be 
the surface of artificial land inside the vessel to allow the whole population to survive with a balanced diet?
How is this dependent on the agricultural technique? We dedicate our third paper of the series to bring a precise 
answer to those questions. To examine such problems, we have improved HERITAGE to include more physical data 
for the simulated crew in order to estimate how much food the settlers need to consume every year. Such investigation
relies on a number of factors such as biological and demographical data, together with their physical activity
(see Sect.~\ref{Improvements}). The total energy expenditure of a given population size can then be translated in terms 
of food quantity to determine the area needed for food production within the spaceship (Sect.~\ref{Space_Farming}).
We examine a variety of scenarios and outcomes, and discuss our results before concluding in Sect.~\ref{Conclusions}.

\section{Improvements to the HERITAGE code}
\label{Improvements}
The HERITAGE code underwent a series of improvements in order to better simulate realistic individuals and populations 
(see the Appendix, Sect.~\ref{Appendix:Inputs}). We now include in the blueprint of each numerical human biological
data on age-dependent pregnancy chances, miscarriage rates and infertility likelihood (see Sect.~\ref{Improvements:biological_data}).
We further added anthropometric data such as the age-dependent height and weight (Sect.~\ref{Improvements:anthropometric_data}).
By doing so it becomes feasible to determine with great precision the total energy expenditure of a stable, heterogeneous
population (Sect.~\ref{Improvements:PAL_TEE}).

\subsection{Biological data on human age-dependent pregnancy, miscarriage and infertility rates}
\label{Improvements:biological_data}

\begin{figure}
\centering
\includegraphics[trim = 0mm 0mm 0mm 0mm, clip, width=8.2cm]{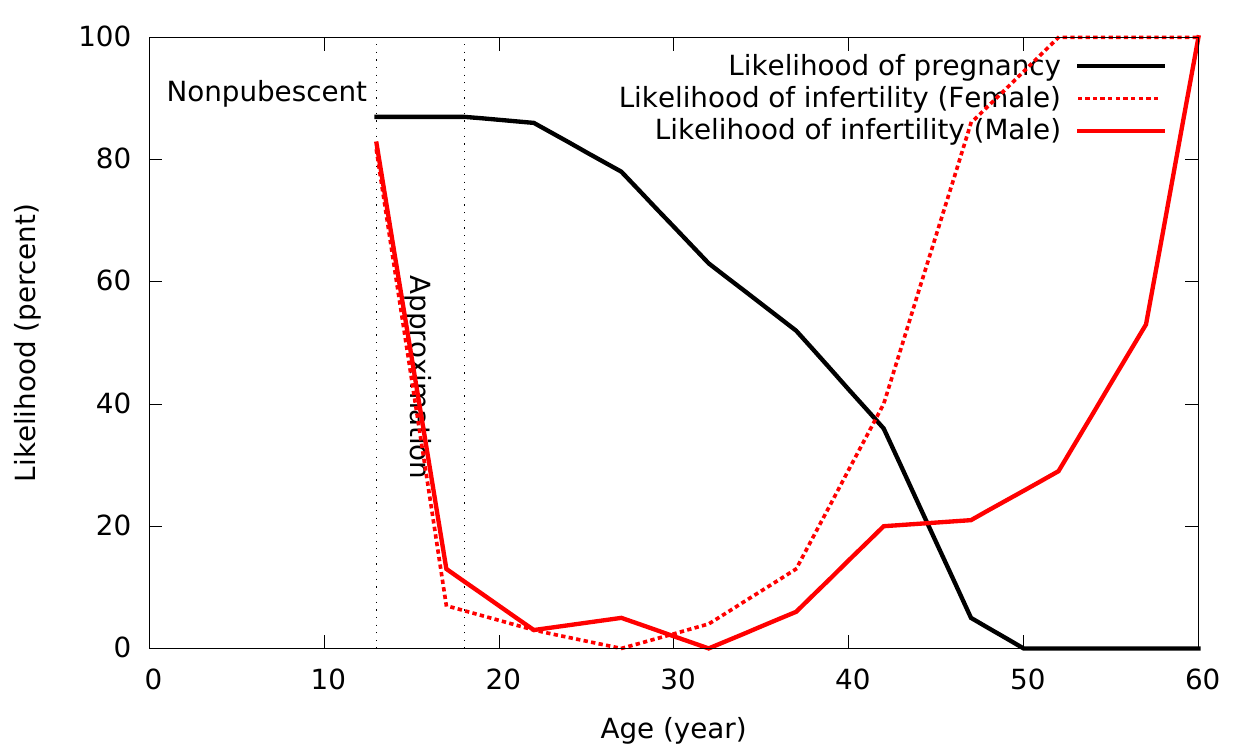}
  \caption{New biological data included in HERITAGE. 
	    Solid black line: pregnancy probability over the 
	    course of one year; solid red line: 
	    age-dependent infertility of a male population;
	    dotted red line: age-dependent infertility 
	    of a female population. Between age 13 and 18
	    the data are only approximative. Before age 13
	    the child is considered as non-pubescent.}
  \label{Fig:Data}
\end{figure}

In the previous version of HERITAGE, the infertility rate and pregnancy chances were age-independent, fixed values.
This was a first-order approximation that we now discard for the benefit of more exact medical data. We include an 
age-dependent pregnancy rate based on reliable etiological sources that discuss the causes of infertility 
and provide information on numerous treatments \cite{Carcio1998,Rosenthal2002}. We also include the male and 
female infertility likelihood based on the work of \cite{Mineau1982}, who explored the age-dependent correlation between 
age and fertility. The resulting age-dependent biological values are presented in Fig.~\ref{Fig:Data}. As we can see the likelihood of pregnancy monotonously 
decreases with time until it reaches zero at age 50. The infertility rates are different between females and males,
with female fertility peaking before 30 while male's fertility peaks right after 30. Both decrease with time 
but female infertility evolves faster and reaches a 100\% infertility rate by 52, while males are fertile for a longer 
period. Between ages 18 and 13 data are incomplete and we can only approximate the global behaviour of the curves. 
Before 13 women are considered as non-pubescent (while this can vary from case-to-case, with extremes down to 10 years 
old \cite{Kail2010}) and for ethical and sociological reasons we do not explore this.

\begin{figure}
\centering
\includegraphics[trim = 0mm 0mm 0mm 0mm, clip, width=8.2cm]{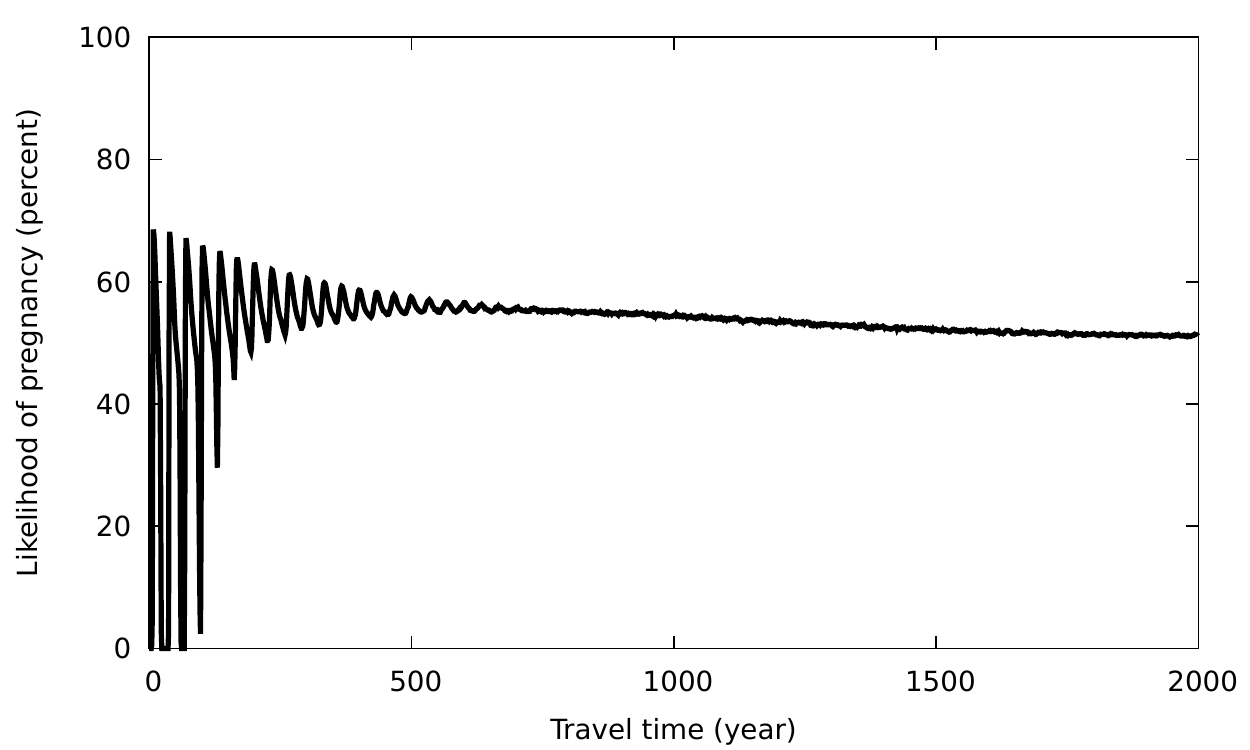}
  \caption{Evolution of the averaged pregnancy 
	    chances per woman considering a 
	    heterogeneous population of several 
	    hundreds by the end of the mission.}
  \label{Fig:Pregnancy_avergared}
\end{figure}     

\begin{figure}
\centering
\includegraphics[trim = 0mm 0mm 0mm 0mm, clip, width=8.2cm]{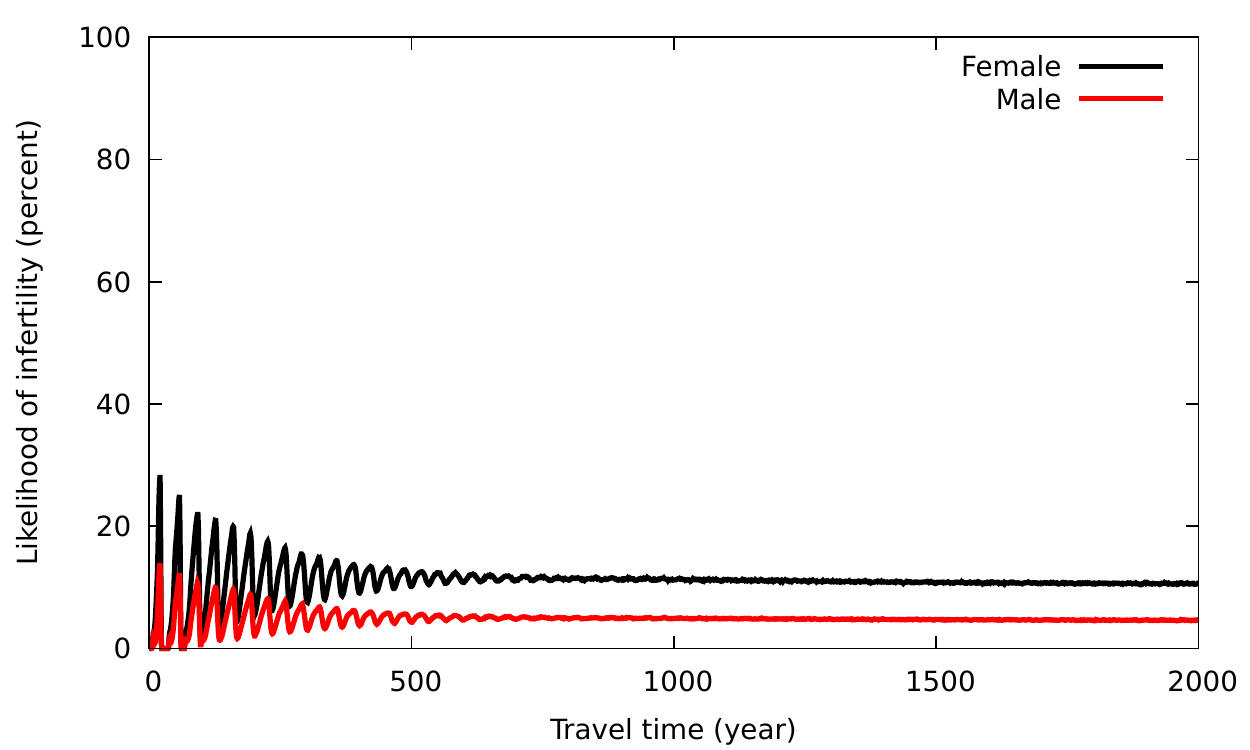}
  \caption{Evolution of the averaged infertility 
	    chances per man (in red) and woman (in black)
	    considering a heterogeneous population of 
	    several hundreds by the end of the mission.}
  \label{Fig:Infertility_avergared}
\end{figure}   

In Fig.~\ref{Fig:Pregnancy_avergared}, we show the evolution of the average pregnancy chance per woman considering 
a heterogeneous population of several hundred by the end of the mission. We ran HERITAGE one thousand times in order to 
get statically significant results (see the Appendix, Sect.~\ref{Appendix:Iterations}). We observe that the global pregnancy
chances are very unstable at the beginning of the interstellar flight due to the presence of well-defined demographic echelons
with people clustering into discrete age groups. When the age groups are too young (non-pubescent) and/or too old (menopause),
their likelihood of pregnancy drops to zero. However, when the population age becomes more evenly spread, a global trend
appears. The population contains about 200 women with different ages, hence different pregnancy chances, and the median value
for pregnancy likelihood stabilizes around 50.5\%. 

In Fig.~\ref{Fig:Infertility_avergared}, we investigate the evolution of the average infertility for males 
and females using the exact same population. Data are statistically noisy at the beginning of the mission for the 
same reasons as presented above, and the curves stabilize when the population becomes stable and heterogeneous in age.
Male infertility is globally lower than female infertility due to slower decrease of fertility with time for men 
(see Fig.~\ref{Fig:Data}). The average infertility within the population is 10.7\% for women and 4.7\% for men 
(plus or minus 0.5\%). Interestingly, we note that our preliminary approximations of age-independent pregnancy rate 
and infertility were valid but only for the case of a stable population. The values we used previously were 75\% for 
pregnancy efficiency, and 10\% and 15\% for female and male infertility respectively. The final values we find 
using age-dependent medical data are lower but compensate each other, in the sense that globally the crew has lower 
pregnancy chances but is more fertile than previously estimated. 

\begin{figure}
\centering
\includegraphics[trim = 0mm 0mm 0mm 0mm, clip, width=8.2cm]{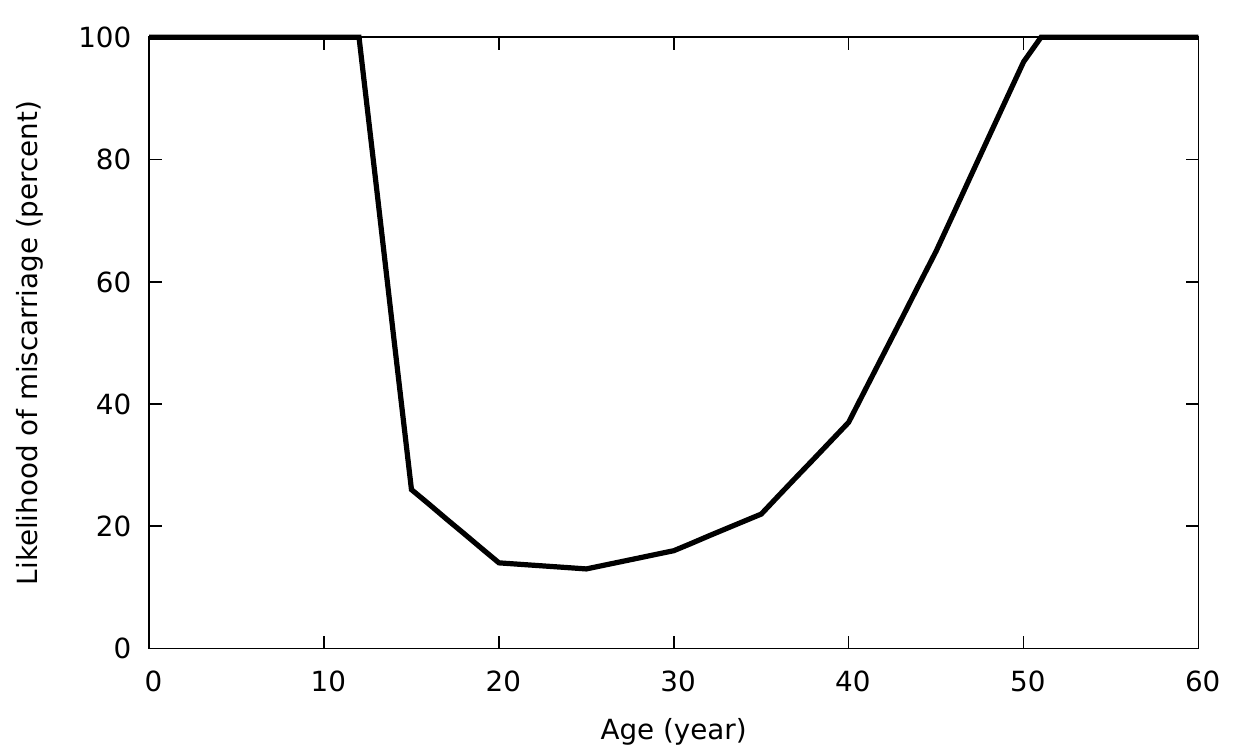}
  \caption{Miscarriage chances as a function 
	  of woman age.}
  \label{Fig:Miscarriage}
\end{figure}    

\begin{figure}
\centering
\includegraphics[trim = 0mm 0mm 0mm 0mm, clip, width=8.2cm]{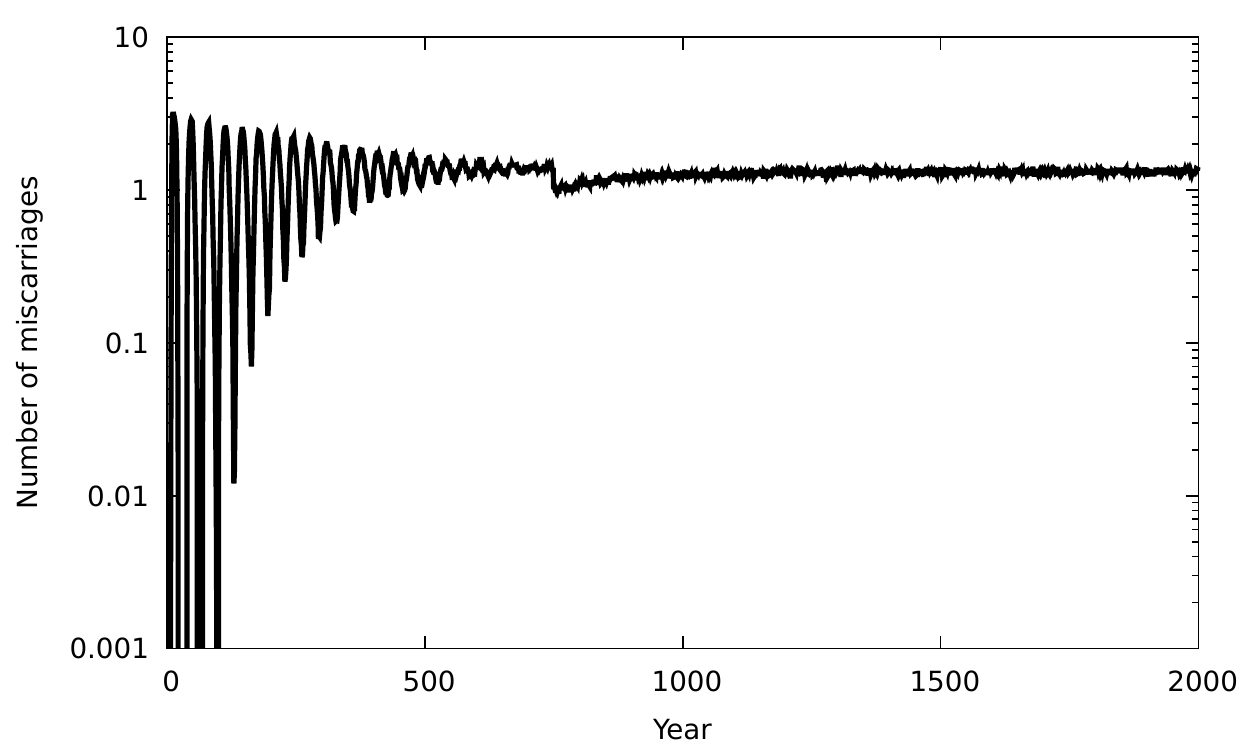}
  \caption{Number of miscarriages per year among the 
	    starship for a 2000 year-long journey. 
	    Data stabilize when the population reaches 
	    a growth stability level. The sharp drop at 
	    750 years is due to a catastrophic event
	    that wipes out 30\% of the population.}
  \label{Fig:Miscarriage_avergared}
\end{figure}   

Another parameter that we now include in our Monte Carlo code is the possibility of miscarriages. Spontaneous 
interruptions of pregnancy often happen before the end of the first trimester of pregnancy and early miscarriages
are mainly caused by a non-hereditary chromosomal abnormality of the embryo (see, e.g., \cite{Rubio2003}). 
Miscarriages are more common for women under 18, in women over 35, and for patients with a history of miscarriage, 
see \cite{Miron2013} and Fig.~\ref{Fig:Miscarriage}. On average, miscarriages occur in about 15\% to 25\% of 
pregnancies \cite{Regan2000,Miron2013}.

Examining the number of miscarriages per year for a 2000 year-long journey in Fig.~\ref{Fig:Miscarriage_avergared},
we see that, similarly to previous plots, results are chaotic before the stabilization of the population that reaches 
$\sim$ 400 humans. On average, there is only one or two miscarriage per year, which represents a very small risk of 
failure for the whole mission. This is due to the fact that sporadic miscarriages are not a disease. Less than 5\% of
women will experience two consecutive miscarriages and less than 1\%, three or more \cite{Regan2000,Miron2013}.
The sharp drop at 750 years is due, as in our previous investigations, to a catastrophic event that wipes out 30\% 
of the population. Note that we restrain this calamity to impact the crew and not the structure of the generation ship, 
so that the integrity of the vessel remains intact.

\subsection{Anthropometric data on human age-dependent height and weight}
\label{Improvements:anthropometric_data}
In order to be more representative of real populations, we decided to include anthropometric data in the initial conditions
of our numerical humans. This has the advantage of being a well documented subject and it is a necessary step to be able 
to calculate the food requirement aboard since the daily caloric consumption depends, among other things, on the age, height and 
weight (see Sect.~\ref{Improvements:PAL_TEE}). We first include the age-dependence of height. To do so, we followed the Dutch 
growth study presented in \cite{Heck2002} to compute the height evolution. The authors used the infancy-childhood-puberty model 
(see \cite{Karlberg1989}) to break down growth mathematically into three partly superimposed components:

\begin{equation}
H_{\rm tot} = H_1 + H_2 + H_3
\label{Height}
\end{equation}

\begin{figure}
\centering
\includegraphics[trim = 0mm 0mm 0mm 0mm, clip, width=8.2cm]{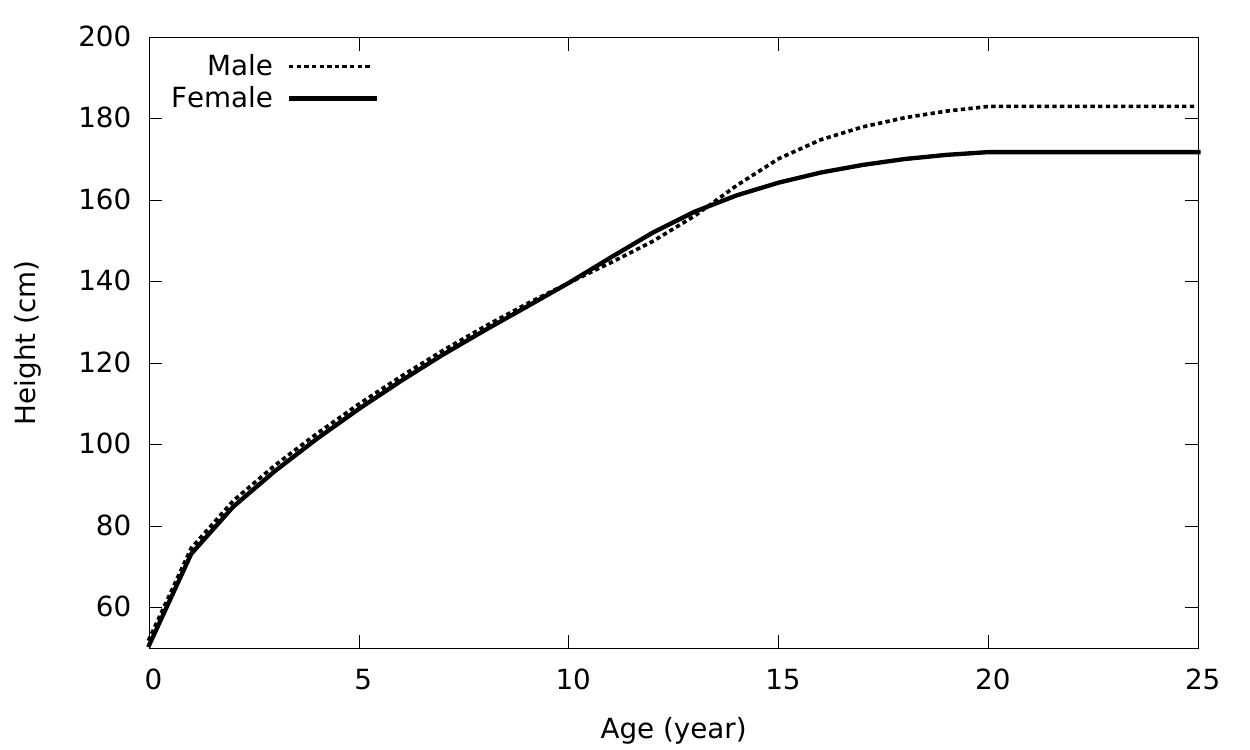}
  \caption{Female (solid black line) and male
	    (dot-dashed black line) stature as 
	    a function of age.}
  \label{Fig:Height}
\end{figure}   

with $H_1$ = 76.4-19.4$e^{-1.56a}$, $H_2$ = -0.235$a^2$+9.5$a$-4.7, $H_3$ = 16.1/(1+$e^{16.4-1.2a}$) for males
and $H_1$ = 74.3-18.7$e^{-1.65a}$, $H_2$ = -0.256$a^2$+9.8$a$-4.8, $H_3$ = 8.6/(1+$e^{12.4-1.1a}$) for females.
In this equation $H_1$ represents the infancy period (0 -- 3 years), $H_2$ the childhood (3 -- puberty) and 
$H_3$ the puberty and onwards. The relative contributions of $H_1$, $H_2$ and $H_3$ strongly vary with $a$, the age in years.
The heights are given in centimeters. The resulting stature $H_{\rm tot}$ as a function of age is shown in
Fig.~\ref{Fig:Height}. We see that, on average, young boys and girls have a similar height until the onset of 
puberty. The final stature of both genders stabilizes around 20 years old, where the curves are plateauing. We
note that the stature of the Dutch population is slightly higher than the averaged height of other populations 
but it is within the worldwide human height interval \cite{Tabassum2009,Moody2013}. Within HERITAGE, we included 
a random 10\% variation of height in order to account for a variety of statures and we checked that the height 
of a given human does not suddenly decrease due to this additional fluctuation. By doing so our population is 
an indiscriminate mix of short, medium and tall people.

Our second anthropometric addition is the inclusion of age-dependent weight. This is slightly more 
tricky as there are numerous methods to estimate body weight. In the case of children there are two 
main equations \cite{Loo2013}. First the Leffler formula (similar to the equations used in the American 
Heart Association training program), which is used for children between 0 and 10 years old:

\begin{equation}
W = 2a+10.
\label{Leffler}
\end{equation}

Then there is the Theron formula that was developed to improve the accuracy of weight estimation for overweight 
children:

\begin{equation}
W = e^{0.175571a+2.197099}
\label{Theron}
\end{equation}

with $W$ the body weight in kilograms and $a$ the age in years. We observe that the weight is the same for both 
male and female children. In the case of adults the situation is different and theoretical curves clearly separate 
men and women. Several equations have been published \cite{Pai2000,Bartlett2012,Peterson2016}. According to the 
Devine formula \cite{Devine1974}, the ideal adult body weight is given by :

\begin{equation}
W_{\rm man} = 50 + 0.9 \times (H - 152) 
\label{Devine_man}
\end{equation}
\begin{equation}
W_{\rm wom} = 45.5 + 0.9 \times (H - 152)
\label{Devine_wom}
\end{equation}

for men and women, respectively. $W$ represents the weight in kg and $H$ the height in cm. Another equation that 
is widely used was presented by Hamwi \cite{Hamwi1964} and takes the form:

\begin{equation}
W_{\rm man} = 48 + 1.1 \times (H - 152) 
\label{Hamwi_man}
\end{equation}
\begin{equation}
W_{\rm wom} = 45.4 + 0.9 \times (H - 152)
\label{Hamwi_wom}
\end{equation}

\begin{figure}
\centering
\includegraphics[trim = 0mm 0mm 0mm 0mm, clip, width=8.2cm]{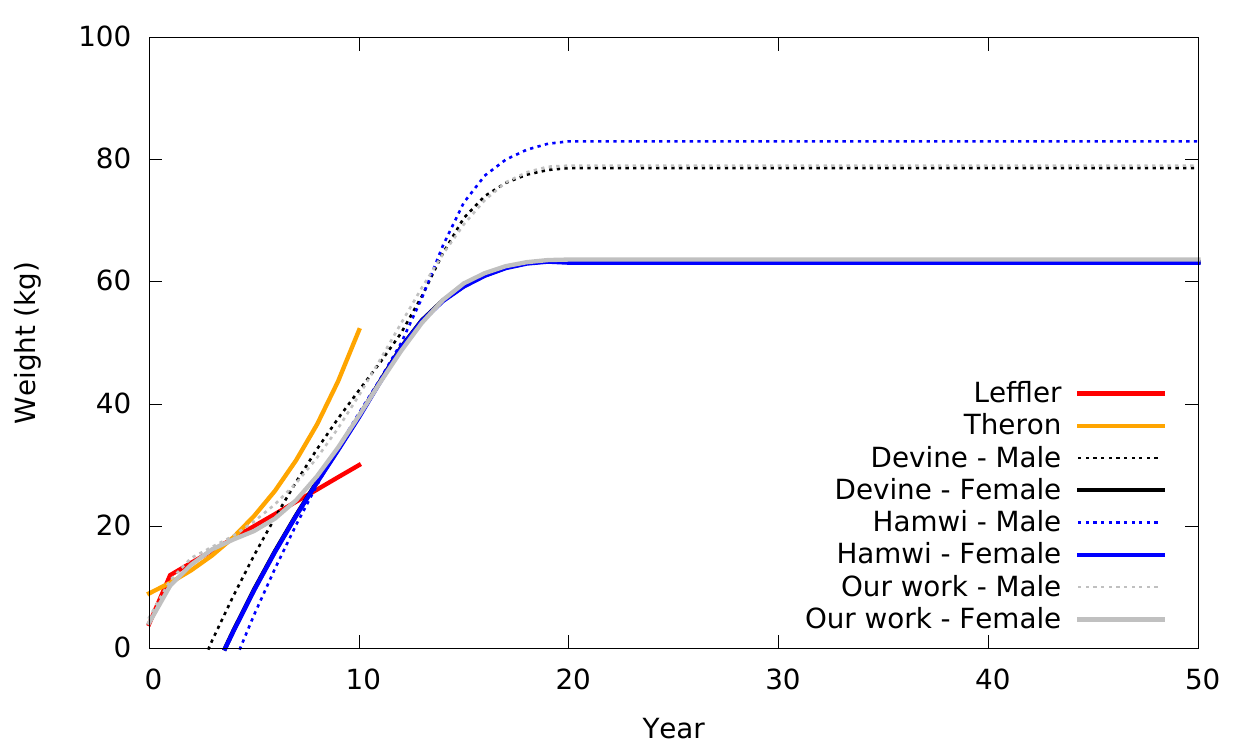}
  \caption{Ideal human body weight according 
	    to various authors (see text for 
	    formulas and references). A
	    fifth order polynomial fit is
	    shown in gray lines.}
  \label{Fig:Weight_formulas}
\end{figure}      

\begin{figure}
\centering
\includegraphics[trim = 0mm 0mm 0mm 0mm, clip, width=8.2cm]{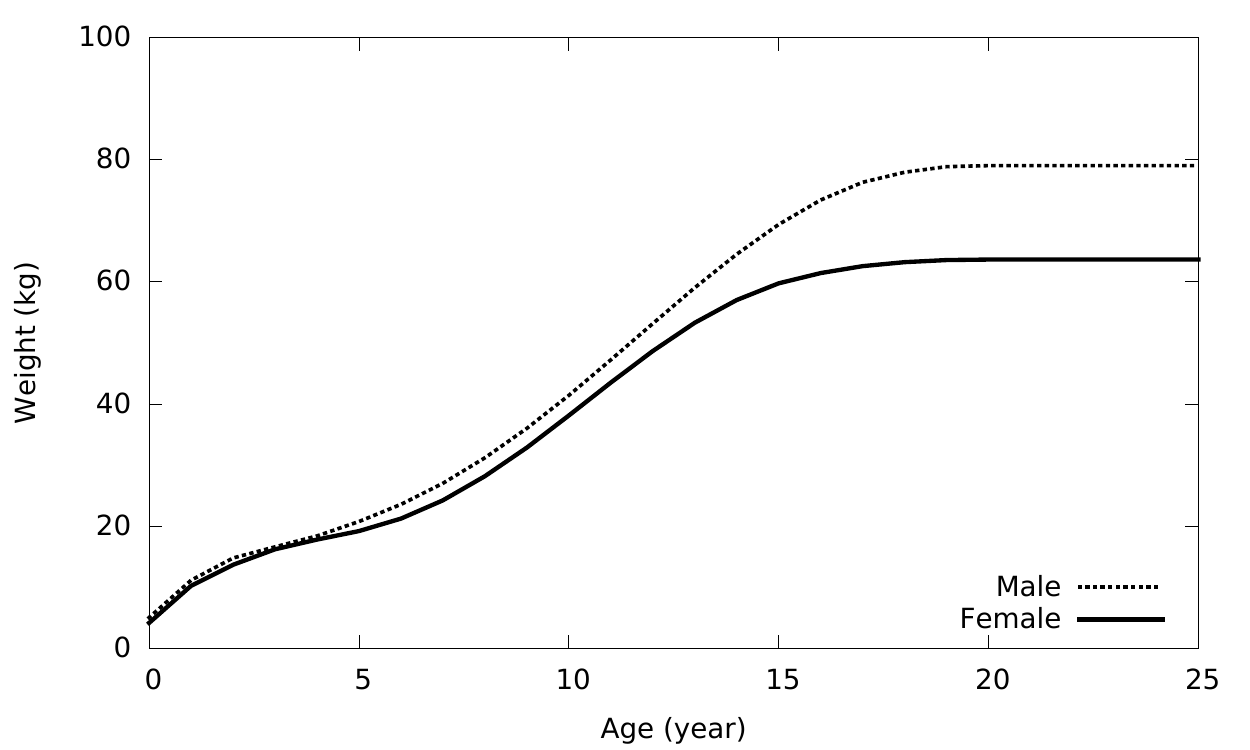}
  \caption{Female (solid black line) and male
	    (dot-dashed black line) weight as 
	    a function of age using a fifth 
	    order polynomial fit from equations
	    presented in Fig.~\ref{Fig:Weight_formulas}.}
  \label{Fig:Weight}
\end{figure}      

Both formulas reach a plateau at age 20 but give quite different results for male weight. In fact, when we plot 
the four different anthropometric equations (Leffler, Theron, Devine and Hamwi, see Fig.~\ref{Fig:Weight_formulas}), 
we find that none of the curves overlap. This is particularly true for the transition between 8 and 12 years.
Since we do not want to rely on a specific equation for HERITAGE, we decided to run a numerical fit to the data.
We extrapolated a fifth order polynomial function that is representative of the four equations. Our formula is 
shown in Fig.~\ref{Fig:Weight_formulas} and with greater details in Fig.~\ref{Fig:Weight}. What is interesting 
with this fit is that males are always heavier than females, even in the childhood period. The difference is 
almost negligible during infancy and increases with time. We thus included this formula in HERITAGE and we took 
into account that, in contrast to height, weight can positively or negatively vary as a function of age. We thus 
allowed the randomly picked weight of each crew member to vary within an interval of 10\% per year so an individual
can be underweight in his childhood and becomes overweight during adulthood. The median value is always centred on our 
fitted formula. This allows the population to have underweight, healthy, and overweight members.

\subsection{Physical activity level and total energy expenditure}
\label{Improvements:PAL_TEE}
Now that we have developed our Monte Carlo code to account for all necessary biological and anthropometric 
data, it becomes feasible to estimate the amount of food needed by the crew during the journey. In order to do so,
we must calculate total energy expenditure. The energy expenditure of the human body 
is mainly a sum of two phenomena: the calories spent by the metabolism to ensure the proper functioning of all 
the vital operations (breathing, digestion, regulation of body temperature) and the calories burnt during physical 
effort associated with external work. The energy consumed by vital organs and the resulting heat 
production is called the basal metabolic rate (BMR). The BMR is strongly correlated with age, height and 
body mass, hence the necessity to include anthropometric formulas in our code. The Harris-Benedict principle
\cite{Harris1918}, revised by \cite{Mifflin1990}, is a method used to estimate an individual's BMR for both
men and women:

\begin{equation}
BMR = (10 \times W_{\rm man}) + (6.25 \times H_{\rm man}) - (5 \times A_{\rm man}) + 5 
\label{Harris-Benedict_man}
\end{equation}
\begin{equation}
BMR = (10 \times W_{\rm wom}) + (6.25 \times H_{\rm wom}) - (5 \times A_{\rm wom}) - 161
\label{Harris-Benedict_wom}
\end{equation}
with $W$ the weight in kg, $H$ the height in cm and $A$ the age in years.

\begin{table*}
  \centering
  \begin{adjustbox}{width=\textwidth}
    \begin{tabular}{lll} 
      \textbf{Lifestyle} & \textbf{Example} & \textbf{PAL}\\
      \hline
      Extremely inactive & Cerebral Palsy patient & $<$ 1.40 \\
      Sedentary & Office worker getting little or no exercise & 1.40 -- 1.69 \\
      Moderately active & Construction worker or person running one hour daily & 1.70 -- 1.99 \\
      Vigorously active & Agricultural worker (non mechanized) or person swimming two hours daily & 2.00 -- 2.40 \\
      Extremely active & Competitive cyclist & $>$ 2.40 \\
      \hline
    \end{tabular}
  \end{adjustbox}
  \caption{Physical activity level (PAL) for several lifestyles.}
  \label{Tab:PAL} 
\end{table*}

\begin{figure}
\centering
\includegraphics[trim = 0mm 0mm 0mm 0mm, clip, width=8.2cm]{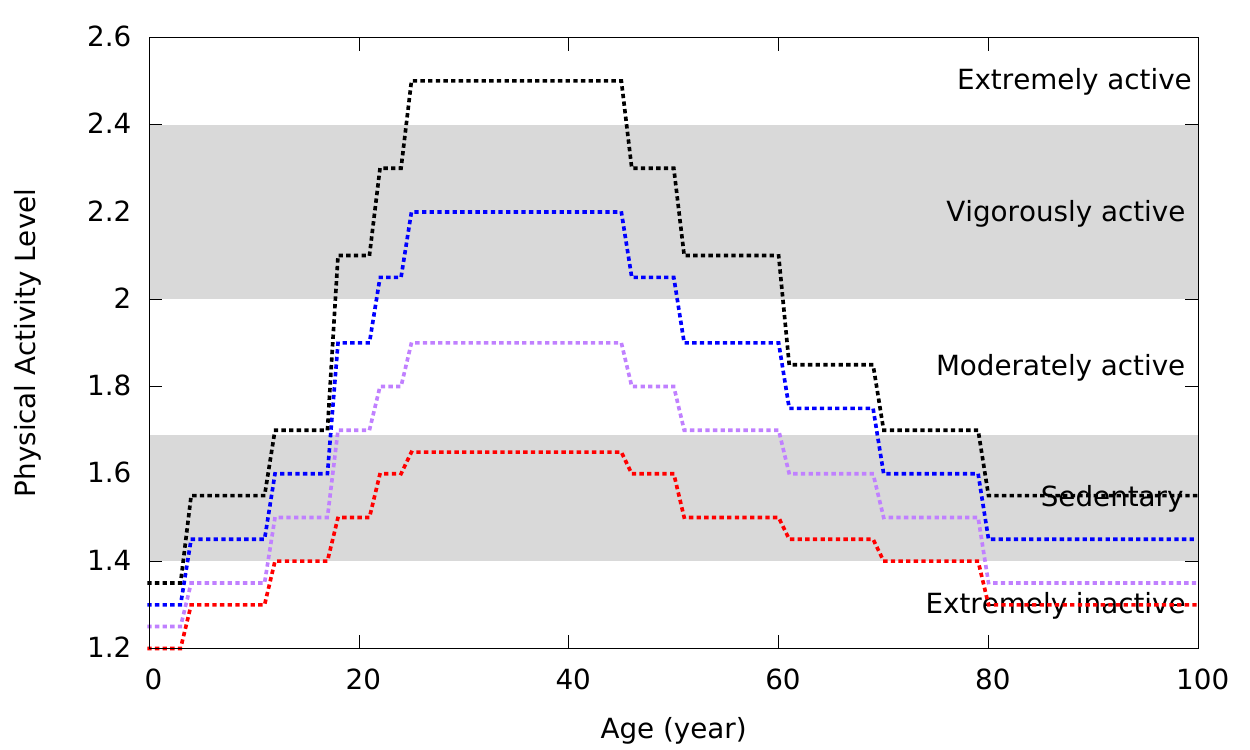}
  \caption{Physical activity level (PAL) 
	    scenarios for four different 
	    populations: sedentary (in red),
	    moderately active (in purple), 
	    vigorously active (in blue),
	    and extremely active (in black).}
  \label{Fig:PAL}
\end{figure}    

The second cause of calorie loss, i.e. the external work, can be evaluated by measuring the physical 
activity level (PAL) of the individual. This is the ratio of energy expenditure to BMR and it can be quantified 
as the sum of obligatory and discretionary physical activities. Obligatory activities include daily activities such as going to school, tending to the home and family,
and other demands made on children and adults by their economic, social and cultural environment, while 
discretionary physical activities are related to health, well-being and quality of life in general.
Five categories of PAL have been identified by the World Health Organization, the United Nations University, and
the Food and Agriculture Organization of the United Nations \cite{PAL2001}. They are summarized in Tab.~\ref{Tab:PAL}.
An extremely inactive PAL corresponds to no exercise at all. A sedentary PAL corresponds to intensive exercise 
for 30 to 60 minutes once to three times a week. This may include activities such as cycling, jogging or swimming, 
or may also corresponds to a busy life style with frequent walks for long periods. A moderately active PAL corresponds 
to intensive exercise for 60 minutes or greater five to seven days a week (same activities as above). Labour-intensive occupations including 
construction work, general labour, farming, or landscape work enter this category. The fourth category, vigorously 
active PAL, corresponds to people with very demanding jobs, such as mining. Finally, the fifth category 
corresponds to exceedingly active and/or very demanding activities such as athletes with an almost unbroken 
training schedule corresponding to multiple training sessions throughout the day.

It is difficult to predict the individual PAL for people living in a starship during their whole life, as it may 
depend on their occupation inside the vessel. Also, some public health rules might be implemented to insure 
everyone has a sufficient activity level to avoid health issues related to obesity \cite{Biswas2015}. In order 
to explore a large parameter phase space, we have considered several scenarios where the individual PAL varies as a 
function of the settler's age. We assume that colonists are not very active in their early years, then there is 
an increase of the activity level that peaks between 25 and 45 years old, and finally the PAL decreases with old 
age. We varied the peak of activity as a function of the scenario we wanted to test, including sedentary, moderately
active, vigorously active, and extremely active populations. The four scenarios are illustrated in Fig.~\ref{Fig:PAL}.

\begin{figure}
\centering
\includegraphics[trim = 0mm 0mm 0mm 0mm, clip, width=8.2cm]{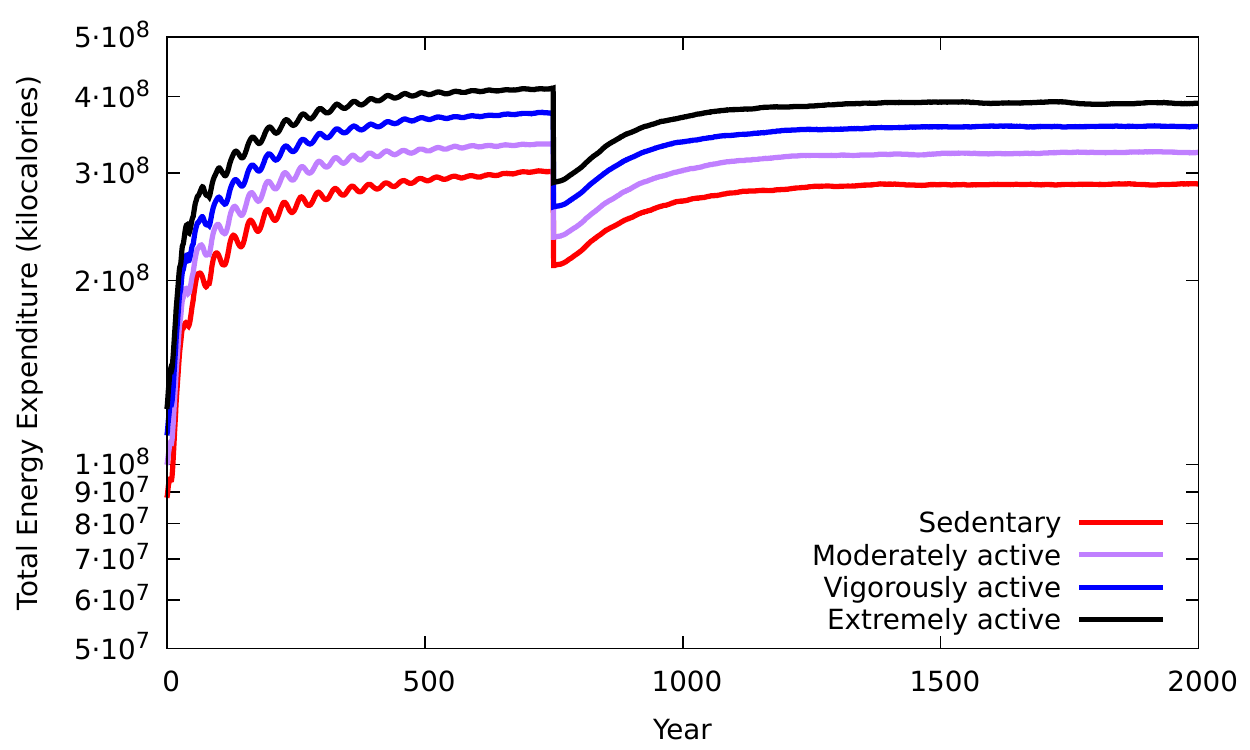}
  \caption{Total energy expenditure (TEE, in kilo-calories) 
	    per year in the vessel for the four different PAL 
	    scenarios presented in Fig.~\ref{Fig:PAL}. The 
	    crew is composed of 98 people at the beginning of 
	    the mission and the population slowly rises to 
	    $\sim$ 400 people by the end. The sharp drop at 
	    750 years is due to a catastrophic event that wipes 
	    out 30\% of the population.}
  \label{Fig:TEE}
\end{figure}  

\begin{figure}
\centering
\includegraphics[trim = 0mm 0mm 0mm 0mm, clip, width=8.2cm]{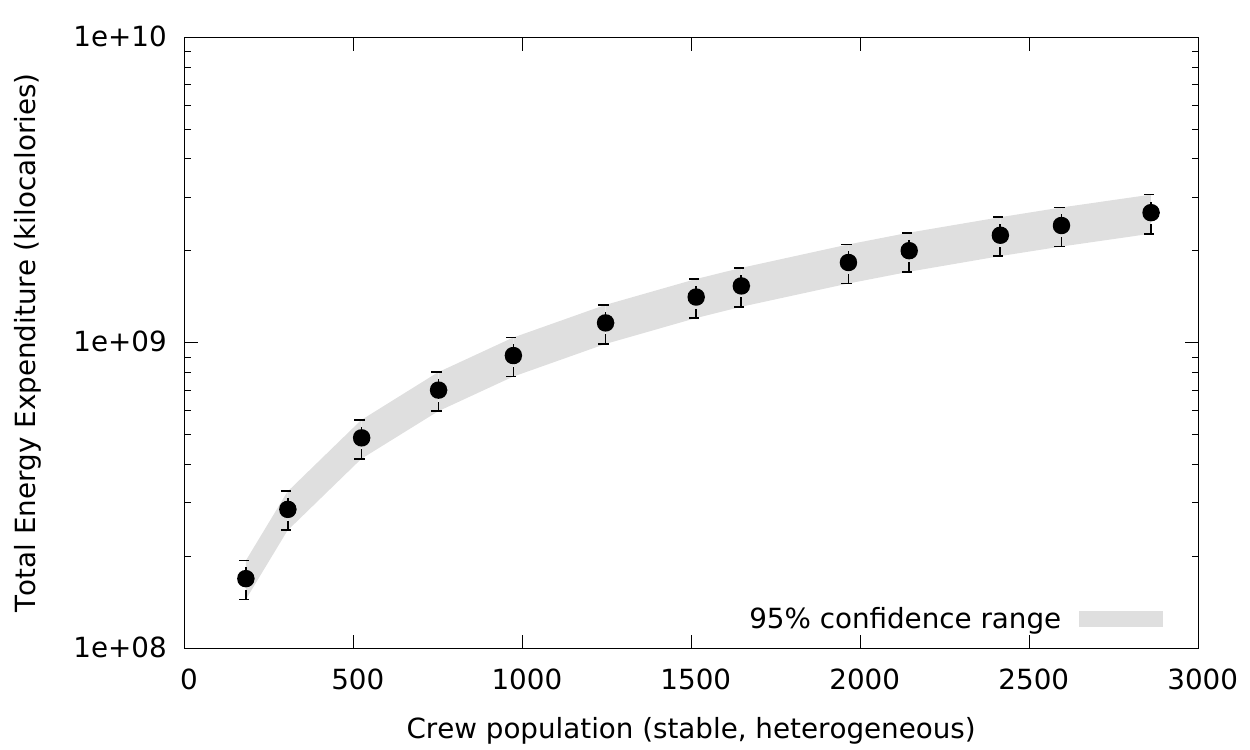}
  \caption{Total energy expenditure (TEE, in kilo-calories)
	    per year as a function of the crew size. The crew 
	    is representative of a vigorously active, stable 
	    and heterogeneous population that includes children, 
	    adults and elders, together with skinny, fat, small
	    or tall persons. The 95\% confidence range is shown
	    in gray.}
  \label{Fig:TEE_vs_Crew}
\end{figure}

The product of BMR and PAL gives the approximate daily kilo-calorie intake needed to maintain current body weight, 
also known as the total energy expenditure (TEE). The TEE has been computed for the four different PAL scenarios presented 
in Fig.~\ref{Fig:PAL} and the results are shown in Fig.~\ref{Fig:TEE}. We find that the total TEE increases as the number of
people in the generation ship increases, from 98 at the beginning of the journey to about 400 people at the end. The population
(and the TEE) stabilizes after 600 years, decreases sharply due to a catastrophic event that wipes out 30\% of the population 
at year 750, then increases again to reach the same maximal level after a few hundred years. It is interesting to note 
that the extremely active population requires only 36\% more calories than the sedentary population. In the extreme 
case where all the 400 settlers are Olympic athletes during their 20's and 30's, one can assume that a maximum of 
4.47 $\times$ 10$^8$ kilo-calories have to be produced every year to correctly feed the whole vessel. This values gives
us an upper limit for food production within the starship. Taking into account a more reasonable yet vigorously active 
population, the total energy expenditure for 400 settlers is about 3.57 $\pm$ 0.52 $\times$ 10$^8$ kilo-calories a year. 
We expended our simulations to smaller and larger populations and plotted in Fig.~\ref{Fig:TEE_vs_Crew} the total 
energy expenditure per year as a function of the crew size, given that the population is both stable and heterogeneous.
The required TEE increases continuously with the population size and can be easily interpolated for larger crews. 

\section{Evaluating the size of artificial land for agriculture}
\label{Space_Farming}
Our simulations of the TEE per year as a function of the crew size enables us to estimate the amount of food required by 
the population, given the constraints of maintaining ideal body weight, avoid cardiovascular risks and sustain a healthy 
lifestyle. Thanks to Fig.~\ref{Fig:TEE_vs_Crew}, we can now estimate the size of artificial land required in the starship 
for agricultural purposes. We proceed in three steps. First we review in Sect.~\ref{Space_Farming:Techniques} the different 
farming techniques that currently exist, including conventional farming, hydroponic and aeroponic methods. Then, to evaluate 
the surface of artificial land to feed any size of population, and finally we apply the different agricultural procedures 
to a single-food diet. By focusing on the highest caloric density food in Sect.~\ref{Space_Farming:Potatoes} we can provide 
a clear lower limit on the agricultural surface. In Sect.~\ref{Space_Farming:Balanced_diet} we examine a more balanced diet
that includes fresh fruits, fresh vegetables, meat, whole grains, nuts, lean proteins and a few other animal-based products 
to be grown/raised aboard.

\subsection{Current and experimental farming methods}
\label{Space_Farming:Techniques}
If we discard agriculture techniques based on an archaic technology with very low productivity, there are essentially
three modern farming techniques. 

The first one is industrial, conventional, geoponic farming. This form of agriculture follows agronomic innovations, uses chemical, 
biological and pesticide fertilizers, uses improved crop varieties and heavy machinery. All of these factors combine to yield
better productivity. Intensive farming, such as needed to feed a population in a starship, comes with serious side effects 
such as soil compaction, soil erosion, and declines in overall soil fertility \cite{Muhammed2018}. To maintain a high productivity,
it will be necessary to let the soil fallow. Fallowing will help the soil to restore organic carbon, nitrogen, phosphorus, and potassium among the principal 
soil nutrients. It was observed that their concentrations significantly increase with increasing fallow duration up to 7 years
\cite{Aguilera2013}. To account for fallowing, it is then necessary to consider a resting period of about a third of the time, 
which means that two artificial land surfaces will produce food while a third surface will lie fallow in order to restore the 
soil organic matter, including water and microbial activity and diversity \cite{Nielsen2011}. To do so, the most efficient method 
is to sow non-edible plant species and use bacterial and fungal organisms to restore the fertility of degraded land \cite{Rashid2016}.

The second farming technique one can consider is hydroponic agriculture. Hydroponics is, in fact, a very old horticultural
technique \cite{Douglas1975} that makes it possible to cultivate a crop above ground. The soil is then replaced by an 
inert and sterile substrate, such as coconut fibers, sand, perlite, coir peat or clay balls. In order to overcome the lack 
of nutrients usually contained in horticultural land, it is necessary to regulate the composition of nutrient solutions using 
automatized engines. An additional improvement with respect to conventional farming is that this technology is not subject to 
weather conditions or the seasons. A study conducted at the John F. Kennedy Space Center by \cite{Mackowiak1989} have shown that it is possible
to grow Triticum aestivum (common wheat) using a growth chamber at 23$^\circ$C, 65\% relative humidity, 1000~ppm CO$_2$, 
continuous light, with a continuous flow, thin film nutrient delivery system. 24 trays of wheat were planted and harvested. 
The grain yields averaged 520~g.m$^{-2}$ and had an average edible biomass of 32\%. More recent studies \cite{Correa2008}
have shown that tuber production from a staggered harvest in hydroponics was 286\% greater than in the bed and pot systems 
for Monalisa and Agata cvs potatoes. This means that, under the hypothesis of functional terrestrial gravity inside the 
spacecraft, hydroponics could improve the food production of cereals, starch, and a variety of fruits and vegetables by almost
300\% with respect to conventional farming. 

In order to get rid of the constraints brought by the use of soil or an aggregate medium, one can explore aeroponic farming
methods. Aeroponics is the process of growing plants suspended in a closed or semi-closed environment by spraying the plant's 
dangling roots and lower stem with an atomized or sprayed, nutrient-rich water solution. The project lead by \cite{Stoner1998}
has shown that this high performance food production technology rapidly grows crops using 99\% less water and 50\% less nutrients
in 45\% less time than geoponic agriculture. The major improvements of aeroponics is that crops can be planted and harvested 
year-round without interruption and it is insensitive to gravity. A more recent aeroponics study of the vegetative growth and 
minituber yield in three potato varieties has shown that the number of minitubers per plant increased by 277.2\% compared 
to hydroponics \cite{Roosta2013}. This found that aeroponic farms are by far more efficient than the other farming techniques,
and the associated technologies are under intense development \cite{Spinoff2006}. 

One major concern is for the production of animal proteins. Hydroponics and aeroponic are capable of growing crops in space
but the space needed to raise a living being cannot be compressed below the limit of the physical size of the animal.
According to the Housing and Space Guidelines for Livestock published by the New Hampshire Department of Agriculture, 
the minimum space requirements for, e.g., a beef or dairy cow is 6.97 -- 9.29~m$^2$. A pig requires a minimum living surface 
of 4.46~m$^2$, a sheep 1.86 -- 2.32~m$^2$, and a turkey 0.56~m$^2$. Those estimates account for decent living conditions, 
not barren battery cages. It is also feasible to dedicate an exercise yard for the animals in order to release both their own and 
the human colonists stress of living in a confined space. The animal that requires the largest exercise area is the horse (18.58~m$^2$).
However, the larger the area, the larger and more complex the spacecraft. In the following we will only account for the
necessary decent living conditions for the required livestock and thus only calculate the agricultural area required on the 
spaceship.

\subsection{First-order approximation: a single-food diet}
\label{Space_Farming:Potatoes}
In order to efficiently feed the population of the vessel, one can consider - as a first-order approximation - a single-food diet.
According to the Food and Agriculture Organization of the United Nations \cite{Oke1990}, a single-food diet based on sweet potatoes 
would be the most efficient hypothesis for our preliminary analysis. For comparison, we report in Tab.~\ref{Tab:Food} the edible 
energy (in kcal/ha/day) for a variety of crops, fruits, vegetables, animals and animal-based products. 

\begin{table}
  \centering
    \begin{tabular}{lll} 
      \textbf{Aliment} & \textbf{Edible energy} & \textbf{Ref.} \\
      \hline
      Sweet potato & 70\,000 & \cite{Oke1990} \\
      Potato & 54\,000 & \cite{Oke1990} \\
      Rice, paddy & 49\,000 & \cite{Oke1990} \\
      Yam & 47\,000 & \cite{Oke1990} \\ 
      Wheat & 40\,000 & \cite{Oke1990} \\
      Groundnut in shell & 36\,000 & \cite{Oke1990} \\
      Cassava & 27\,000 & \cite{Oke1990} \\
      Lentil & 23\,000 & \cite{Oke1990} \\
      Carrots & 20\,500 & \cite{Crane1951} \\
      Milk & 17\,400 & \cite{Cooper1917} \\     
      Orchard & 16\,900 & \cite{Graham2017} \\      
      Meat: pork & 16\,500 & \cite{Cooper1917} \\      
      Cheese & 10\,400 & \cite{Cooper1917} \\
      Butterfat & 8\,700 & \cite{Cooper1917} \\     
      Meat: mutton & 3\,300 & \cite{Cooper1917} \\
      Meat: beef & 3\,200 & \cite{Cooper1917} \\
      Eggs\tablefootnote{\label{1sttablefoot}Assuming 
	   that poultry are kept under ordinary 
	   poultry farm conditions, the pullets 
	   being raised and the old hens and 
	   young males being used for meat.} & 1\,900 & \cite{Cooper1917} \\    
      Meat: poultry\footref{1sttablefoot} & 1\,700 & \cite{Cooper1917} \\    
      Honey & 1\,400 & \cite{Crane1951} \\      
      \hline
    \end{tabular}
  \caption{Average energy production of selected 
	   aliments in kcal/ha/day.}
  \label{Tab:Food}
\end{table}

\begin{figure}
\centering
\includegraphics[trim = 0mm 0mm 0mm 0mm, clip, width=8.2cm]{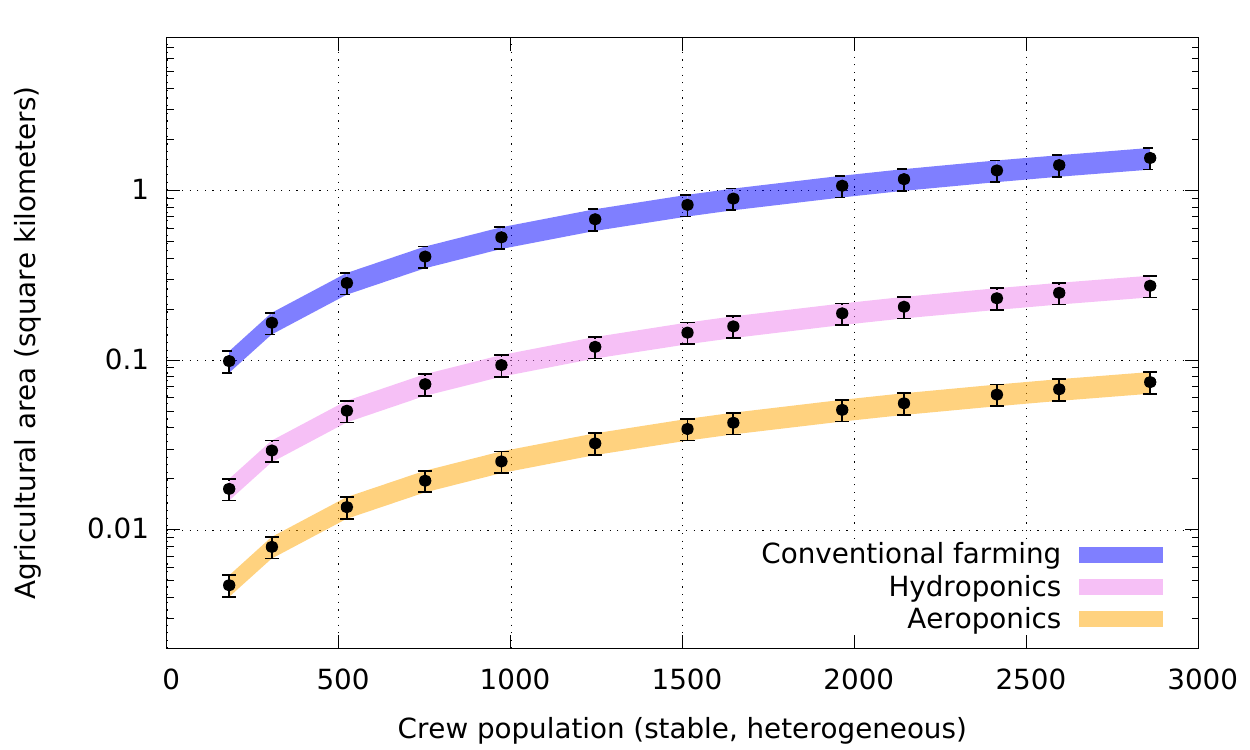}
  \caption{Required agricultural surface area (in 
	    square kilometers) as a function of the 
	    crew size for a single-food diet, see 
	    Sect.~\ref{Space_Farming:Potatoes}.
	    The colors highlight the different,
	    farming techniques used.}
  \label{Fig:Sweet_potatio_diet}
\end{figure}   

We see in Fig.~\ref{Fig:Sweet_potatio_diet} the required agricultural surface area (in square kilometers) as 
a function of the crew size for a sweet potato diet. Using conventional farming techniques, a crew of 500 
would require an agricultural area of 0.230 -- 0.315~km$^2$ to grow food. Using hydroponics, this surface is reduced 
to 0.042 -- 0.055~km$^2$. With aeroponics farms, growing enough sweet potatoes would only require 0.012 -- 0.015~km$^2$.

\subsection{Second-order approximation: a balanced diet}
\label{Space_Farming:Balanced_diet}
Our previous results are a useful first step, but a society eating nothing (or almost nothing) but sweet potatoes 
would likely be riddled with disease. In order to overcome deficiencies in proteins, vitamins, carbohydrates, 
fat, iron, calcium and other elements, it is necessary to construct a balanced diet. To achieve this goal we follow 
the dietary advice from Public Health England \cite{PHE2016}. They recommend to eat each day a selection of fruits
and vegetables (39\%), meat and fish (12\%), dairy (8\%) and starch (37\%) at the proportions given in parentheses.
About 1\% of the dietary content should be used for oils and spreads and 3\% in food high in salt, sugar 
and fat. Based on Tab.~\ref{Tab:Food} and on the constraints established in Sect.~\ref{Space_Farming:Techniques},
it is possible to evaluate the average energy production of each food category to sustain a healthy diet.

\begin{figure}
\centering
\includegraphics[trim = 0mm 0mm 0mm 0mm, clip, width=8.2cm]{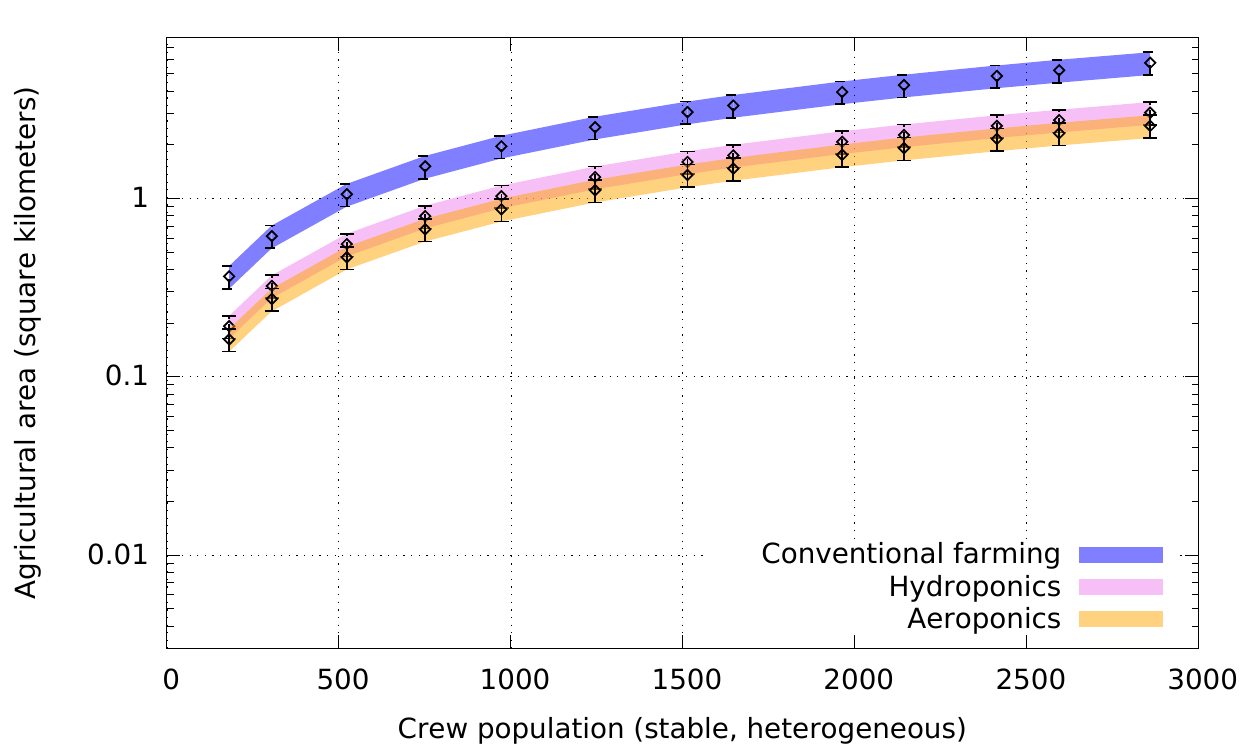}
  \caption{Required agricultural surface area (in 
	    square kilometers) as a function of the 
	    crew size for a balanced diet, see 
	    Sect.~\ref{Space_Farming:Balanced_diet}.
	    The colors highlight the different
	    farming techniques used.}
  \label{Fig:Balanced_diet}
\end{figure}    

In Fig.~\ref{Fig:Balanced_diet}, we computed the required agricultural surface area as a function of the 
crew size for a balanced diet that comprises fruits, vegetables, meat, fish, dairy, and starch. As stated 
previously, the animals that are used for proteins, dairy or honey are kept under decent living conditions 
with enough space to move. We see that, overall, the space needed to fulfill a balanced diet has drastically 
increased with respect to a single-food diet. This is because the average edible energy per product is lower 
than for sweet potatoes. For a crew of 500, the different farming techniques require, on average, a surface
of 1.01~km$^2$, 0.53~km$^2$, and 0.45~km$^2$, for conventional farming, hydroponics and aeroponics respectively.
The 1.01~km$^2$ geoponic value (i.e. 0.2 of a hectare per person) compares very well with the minimum amount of 
agricultural land necessary for sustainable food security, with a diversified diet similar to those of North America
and Western Europe (hence including meat), that is 0.5 of a hectare per person according to the Food and Agriculture
Organization of the United Nations \cite{FAO1993}. We see that the decrease in required land area due to hydroponics
is only a factor 2 with respect to conventional farming. The reduction factor is even lower between hydroponics and
aeroponics because of the physical constraints brought by the animals. Successful further improvement in crop 
production would not help to decrease the size of artificial land needed for agriculture by a large factor because
of the presence of animal proteins and dairy in a balanced diet. The presence of animals on board the spacecraft 
determines the limit of the land size required for agriculture. We will discuss alternative methods for protein 
production in Sect.~\ref{Conclusions}.

\subsection{Ship's architecture}
\label{Space_Farming:Ship}
We saw that a minimum of 0.45~km$^2$ of farmland, using a mix of aeroponics (for fruits, vegetables, starch, 
sugar, and oil) and conventional farming (for meat, fish, dairy, and honey), is to be allocated in the spaceship. 
While aeroponics is insensitive to gravity, animals and humans suffer deleterious health effects from long periods
without (Earth-like) gravity - for example bone demineralization, with bone density dropping at over 1\% per month 
(about 12 times faster than for elderly people on Earth \cite{White2001}). Micro-gravity also influences muscles, 
the heart and brain, and increases cancer risk. To prevent such problems, some form of Earth-like gravity is necessary.
Artificial gravity can be created using a centripetal force \cite{Clement2015}. In this case, the resulting `gravity'
is the inertial reaction to the centripetal acceleration that acts on a body in circular motion. There are four 
interdependent parameters that are used to constraint the system: the radius $R$ from the center of rotation, the 
angular velocity $\Omega$, the tangential velocity $V$, and the centripetal acceleration $A$. 

The radius $R$ corresponds to the distance from the centre of rotation. A natural geometry for the spaceship 
is a cylinder that is rotating like a rigid body. Since the nominal artificial gravity is directly proportional 
to $R$, inhabitants will experience a head-to-foot gravity gradient. To minimize this, one should 
maximize the radius. 

The angular velocity $\Omega$ is the rate at which the spaceship rotates around its centre. The cross-coupling of normal 
head movements (i.e. rotation) with the habitat rotation can lead to dizziness and motion sickness, so, to minimize this 
cross-coupling, one should minimize the habitat's angular velocity. A threshold of $\Omega \le 3$ is recommended to avoid 
disabling motion sickness \cite{Graybiel1977}.
 
The tangential velocity $V$ is the velocity measured at any point tangent to the spaceship rotating surface. As humans 
will move within a rotating habitat, they will be subject to Coriolis accelerations that distort the apparent gravity. 
For relative motion in the plane of rotation, the ratio of Coriolis to centripetal acceleration is twice the ratio of 
the relative velocity to the habitat's tangential velocity. This ratio can be minimized by maximizing the the habitat's
tangential velocity. 

Finally the centripetal acceleration $A$ (measured in units of gravitational force $g$), is a measurement of the 
type of acceleration that causes a perception of weight. For example, 1~$g$ is the acceleration due to gravity at 
the Earth's surface. While the minimum $g$ required to preserve health remains unknown, the maximum $A$ should 
generally not exceed 1~$g$ for comfort reasons \cite{Stone1973}. A safe range of $g$-values lies between 0.3 and 1 
according to \cite{Stone1973,Gilruth1969,Gordon1969}.

\begin{table*}
  \centering
    \begin{tabular}{ccccc} 
      \textbf{$A$ (g)} & \textbf{$\Omega$ (rotations/min.)} & \textbf{$V$ (m/s)} & \textbf{$R$ (m)} & \textbf{$L$ (m)} \\
      \hline
      1.0 & 1.998 & 46.87 & 224 & 320 \\  
      0.9 & 1.996 & 42.22 & 202 & 355 \\  
      0.8 & 1.999 & 37.47 & 179 & 400 \\  
      0.7 & 1.997 & 32.83 & 157 & 456 \\  
      0.6 & 1.994 & 28.18 & 135 & 531 \\  
      0.5 & 1.998 & 23.43 & 112 & 639 \\      
      0.4 & 1.994 & 18.79 & 90 & 796 \\        
      0.3 & 1.986 & 14.14 & 68 & 1053 \\              
      \hline
    \end{tabular}
  \caption{Ship's architecture for a crew of 500 humans
	   and the associated 0.45~km$^2$ of farmland
	   needed to feed them. The length $L$ of the
	   starship is calculated from the usual 
	   surface area of a cylinder: 2$\pi R L$.
	   The ship's radius and length do not account 
	   for other facilities besides farming.}
  \label{Tab:Ship}
\end{table*}

Knowing the surface of farmland in the spaceship, it is then possible to calculate possible ship architectures
under the hypothesis of centripetal gravity. To do so, we used the open-access spin calculator \textit{SpinCalc}
developed by Theodore W. Hall. We present in Tab.~\ref{Tab:Ship} a representative set of spaceship 
radii $R$ (and their associated length $L$), in meters, to maintain a surface of 0.45~km$^2$ of farmland
under different centripetal accelerations. We varied the values of $A$ and maintained $\Omega$ close to 2. 
We see that, for decreasing artificial gravity, the tangential velocity $V$ and $R$ decrease, resulting in 
increasing values of $L$. A cylindrical generation ship with an Earth-like gravity could have a radius
of 224~m and a length of 320~m in order to maintain a surface of 0.45~km$^2$ of farmland, whilst minimizing the 
various uncomfortable effects of rotational gravity. The length of the spaceship is, of course, a simple approximation. 
It seems reasonable to assume that plant and food crops may tolerate a lower level of gravity. Thus if we allow the 
ship to contain multiple floors, each with a different radius, the required area can be maintained while significantly
reducing the length of the cylinder. Assuming the depth of each level to be 3\,m, if we allow food crops on
levels down to 0.9\,$g$, the length of the 224\,m-radius cylinder could be reduced to 106\,m, or 25\,m if we allow them down
to 0.5\,$g$. Of course other facilities besides farming are necessary - human habitation, control rooms, power generation,
reaction mass and engines, the requirements of which we leave to future papers.

\section{Conclusions and further development}
\label{Conclusions}
In this paper, we have improved our Monte Carlo code in order to account for all the necessary biological and anthropometric 
data to compute the yearly energy expenditure aboard a multi-generational spacecraft. We tested a complete phase space of 
scenarios that simulated different crew activity levels that directly impact the caloric budget. By doing so, we determined 
the required amount of kilo-calories to be consumed per year in the spaceship in order to maintain the ideal body weight of 
a stable, heterogeneous crew. The final relation between the yearly kilo-calorie expenditure and the size of the crew is
established. This allows us to easily determine the amount of food to be produced aboard. We investigated the problem through 
the prism of different farming techniques: conventional agriculture, hydroponic farms, and aeroponic systems. The latest 
is the most efficient method to harvest large quantities of crops with minimum space requirement. It also works under low
gravity conditions and does not require soil. We used those three techniques to determine the minimum size of artificial 
land to be saved in the spacecraft for agricultural purposes. First considering a single-food diet, we determined that 
aeroponic farms could produce enough food to feed a population of 500 with only 0.012~km$^2$ of farming area. Improving the diet to include 
dairy, meat and a large variety of starch, vegetables and fruits, the final land surface is larger. An area of 0.45~km$^2$ 
is required to grow all the food, but also to raise animals in decent living conditions for protein and dairy production. 
Working with the hypothesis of centripetal artificial gravity, this surface put strong constraints on the ship's architecture.

We have seen in Sect.~\ref{Space_Farming:Balanced_diet} that the limiting factor to the agricultural surface 
is the presence of animals. Hydroponics and aeroponics are able to reduce the size of vegetable crop cultures but reducing the space associated with
protein intake is a much more complicated task. However potential solutions exist. In particular, there is a growing 
research area on edible insects as an alternative protein source for human food and animal feed. Insects
are highly nutritious and thus represent an interesting alternative to animal meat. A list of insects to be
used in gastronomy is presented in \cite{Rumpold2013}, together with a discussion on the risks and benefits of
insects as a human food source. The harvesting of insects is also much simpler than for cows or poultry, and it requires much 
less space. According to \cite{Hanboonsong2013}, an average of 7500 metric tonnes of insects is annually 
produced for home consumption and markets in Thailand. In fact, roasted mealworms have a higher protein content 
than chicken, pork or beef \cite{Jones2015}. The author of the aforementioned paper calculated that about 160,000 
mealworms per day must be eaten by a gender-balanced crew of 160 people in order to fulfil their daily protein 
requirement. This, of course, opens a whole new window as it could help to decrease the size of artificial land 
used in the spacecraft, but insects can also be used to improve biodiversity, regenerate soils, and destroy organic
wastes while providing extra fertilizers. Insect consumption, however, is not frequent in all countries and a 
psychological barrier must be crossed.

An important and still untouched question regards water. The National Academy of Medicine suggests that an average 
male adult should drink 3.7 litres of water daily, while an average female adult should drink 2.7 liters per day 
in order to maintain body functions \cite{Gibson2012}. Of course, water needs vary tremendously from individual to 
individual, and are dependent on numerous factors such as the activity level or the environmental temperature. Most 
people will be adequately hydrated at levels well below these recommended volumes. It must be noted that these amounts
include water from food consumption. About 20\% of the daily total water intake can be found in food \cite{Nielsen2010}.
If we consider a gender-balanced crew of 500 persons, approximately 468,000 litres of water will be required every year. 
This represents a storage of volume 468 m$^3$ and, apart from this volume, one also needs to take into account the 
mass of the containers. 

Since water refuelling will not be possible during the journey, a recycling system should be 
installed in the spaceship. Such system is currently used aboard the ISS. Astronaut's waste water is captured, such 
as urine, sweat, and even the moisture from their breath. Then impurities and contaminants are filtered out. The final 
product is potable water which is cleaner than what most Earthlings drink. However, the system is not 100\% efficient: 
water is lost by the space station in several ways: a small amount of urine cannot be purified; the oxygen-generating
system consumes water; air that is lost in the air locks takes humidity with it; the CO$_2$ removal 
systems leach some water out of the air, to name a few \cite{Carter2012}. So the amount of water to be stored at the 
beginning of the journey has to take into account the water recycling systems efficiency in order to have enough water
for all the colonists inside the vessel during the whole trip. 

Finally, estimating the amount of water required for plant growth is extremely complex since it is species-dependent. It 
is impossible, as this stage, to give a precise number of how much water should be embarked, but this number is expected 
to be very large. Clearly, the best but difficult to resolve option would rely on finding alternative water sources along 
the ship's journey, e.g. from comets, asteroids, and other large sources while still in the Solar System. Similarly,
the issue of how efficiently plant and animal nutrients can be recycled will strongly affect the amount of mass required for
sustainable farming aboard the ship.

In conclusion, we have put strong constraints on the morphological structure of future multi-generational spacecraft. 
By improving our simulation tool, we have opened a new field of investigation for HERITAGE. There are still many 
more steps to be taken in order to provide a realistic simulation of a global generation ship and we aim at 
pushing our numerical tool to higher grounds by including population genetics and mutation in the next paper of this series.

\section*{Acknowledgment}
The authors would like to thank Theodore W. Hall for his JavaScript applet ``SpinCalc'' 
(https://www.artificial-gravity.com/sw/SpinCalc/) that was used to calculate the various artificial-gravity environments.
R. Taylor was supported by the Albert Einstein Centre for Gravitation and Astrophysics via the Czech Science Foundation 
project 14-37086G, the institutional project RVO 67985815, and the Czech Ministry for Education, Youth and Sports 
research infrastructure grant LM 2015067.

\bibliography{mybibfile}

\newpage

\appendix

\section{Input parameters for HERITAGE}
\label{Appendix:Inputs}
In Tab.~\ref{Tab:Parameters}, we remind the reader about the list of parameters that must be determined by the user 
before starting the simulation. Extensive explication, details and description of the parameters are given in 
\cite{Marin2017} and \cite{Marin2018}.

\begin{table*}
  \centering
    \begin{tabular}{lrl}
	\textbf{Parameters} & \textbf{Values} & \textbf{Units}  \\
        \hline
        Number of space voyages to simulate & 1000 & -- \\
	Duration of the interstellar travel & 2000 & years \\
	Colony ship capacity & 500 & humans \\
	Overpopulation threshold & 0.9 & fraction \\	
	Inclusion of Adaptive Social Engineering Principles (0 = no, 1 = yes) & 1 & -- \\	
	Number of initial women & 49 & humans \\
	Number of initial men & 49 & humans \\
	Age of the initial women & 20 & years \\
	Standard deviation for the age of the initial women & 1 & years \\
	Age of the initial men & 20 & years \\
	Standard deviation for the age of the initial men & 1 & years \\
	Number of child per woman & 2 & human \\
	Standard deviation for the number of child per woman & 0.5 & human \\
	Twinning rate & 0.015 & fraction \\
	Life expectancy for women & 85 & years \\
	Standard deviation for women life expectancy & 15 & years \\
	Life expectancy for men & 79 & years \\
	Standard deviation for men life expectancy & 15 & years \\
	Mean age of menopause & 45 & years \\
	Start of permitted procreation & 30 & years \\
	End of permitted procreation & 40 & years \\
	Initial consanguinity & 0 & fraction \\
	Allowed consanguinity & 0 & fraction \\
	Life reduction due to consanguinity & 0.5 & fraction \\	
	Possibility of a catastrophic event (0 = no, 1 = yes) & 1 & -- \\
	Fraction of the crew affected by the catastrophe & 0.3 & fraction \\
	Year at which the disaster will happen (year; 0 = random) & 750 & years \\
	Chaotic element of any human expedition & 0.001 & fraction \\
        \hline
    \end{tabular}
    \caption{Input parameters of the simulation.}
    \label{Tab:Parameters}
\end{table*}

\section{How many iterations needed?}
\label{Appendix:Iterations}
All simulations presented in this paper have been achieved by looping our Monte Carlo code one thousand times.
The reader may wonder if this number is sufficient to have statistically significant results. 

\begin{figure*}
    \centering
    \begin{subfigure}[b]{0.475\textwidth}
	\centering
	\includegraphics[width=\textwidth]{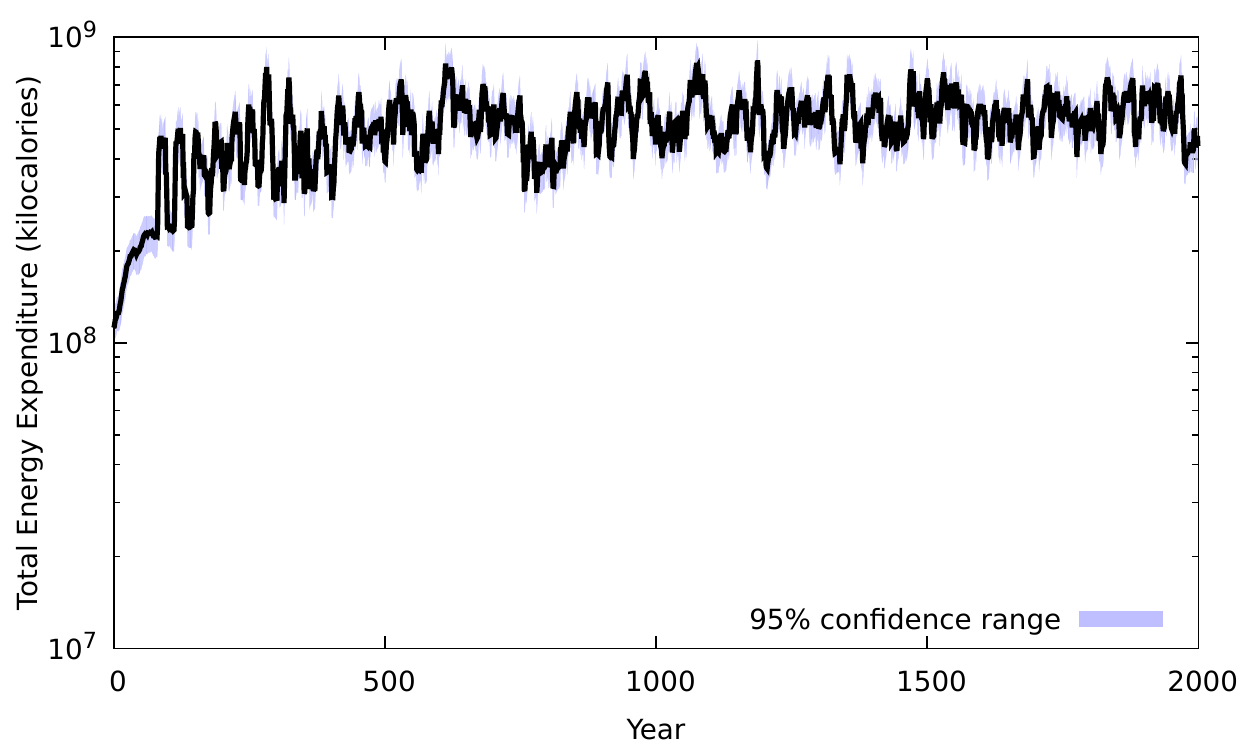}
	\caption{{\small 1 interstellar trip sampled (1 iteration).}}    
    \end{subfigure}
    \hfill
    \begin{subfigure}[b]{0.475\textwidth}  
	\centering 
	\includegraphics[width=\textwidth]{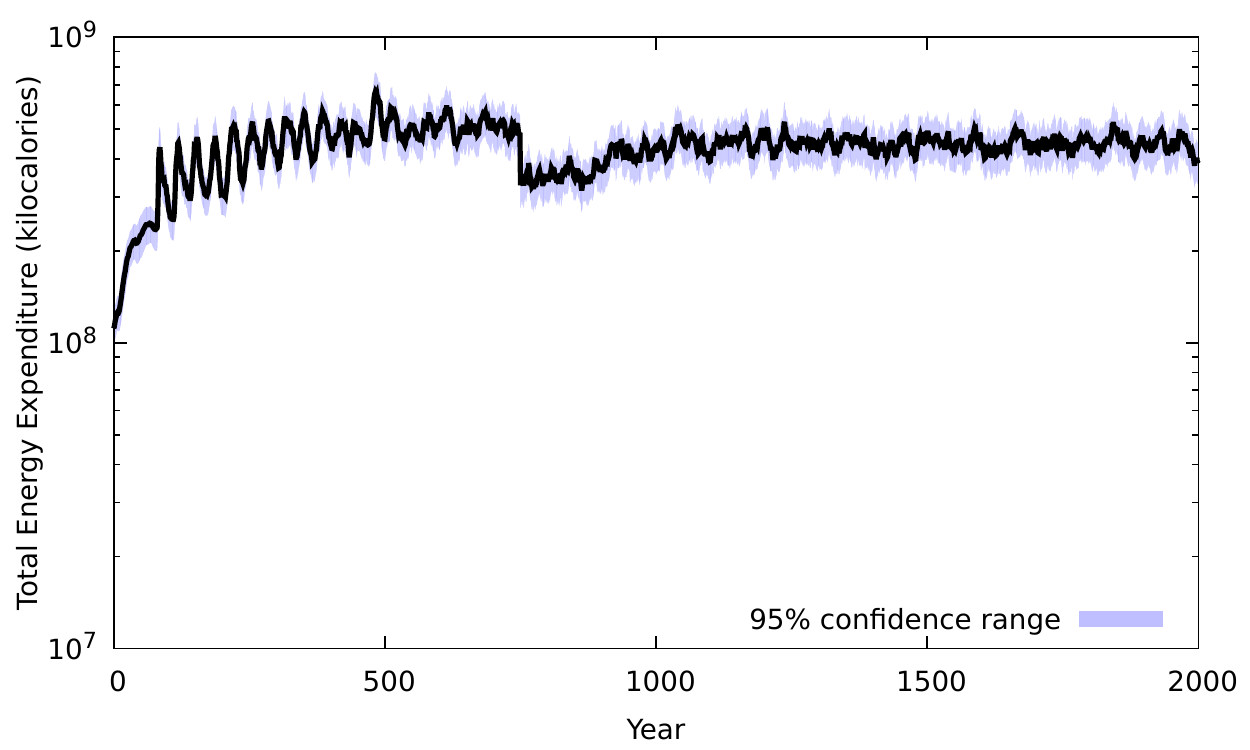}
	\caption{{\small 10 interstellar trips sampled (10 iterations)}}    
    \end{subfigure}
    \vskip\baselineskip
    \begin{subfigure}[b]{0.475\textwidth}   
	\centering 
	\includegraphics[width=\textwidth]{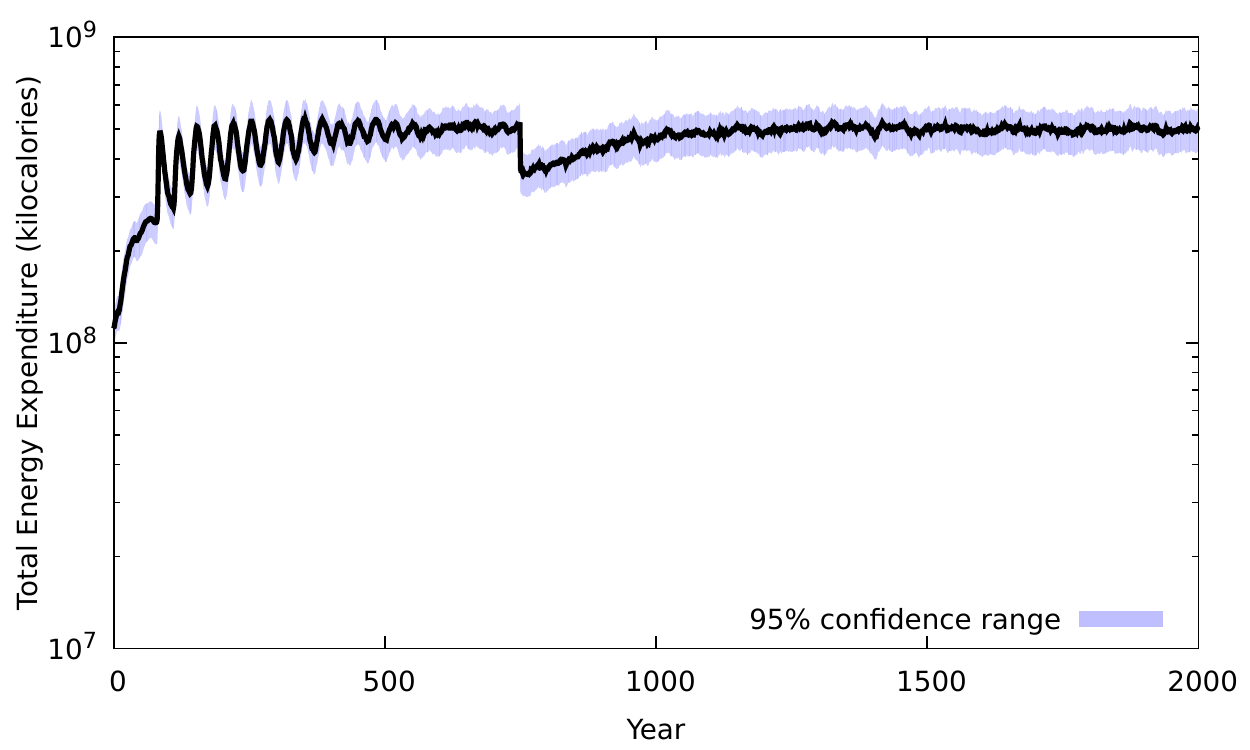}
	\caption{{\small 100 interstellar trips sampled (100 iterations)}}    
    \end{subfigure}
    \quad
    \begin{subfigure}[b]{0.475\textwidth}   
	\centering 
	\includegraphics[width=\textwidth]{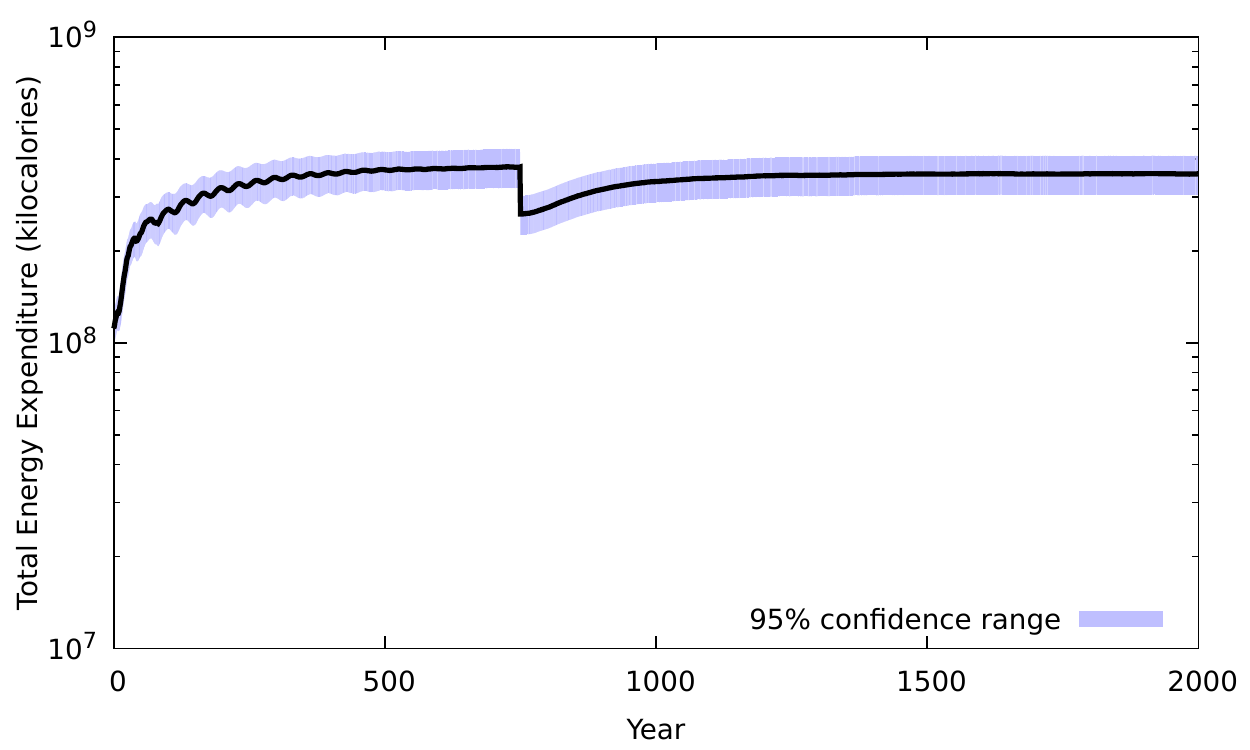}
	\caption{{\small 1000 interstellar trips sampled (1000 iterations)}}    
    \end{subfigure}
    \caption{Total energy expenditure (TEE, in kilo-calories) 
	    per year in the vessel for a vigorously active population
	    (see Figs.~\ref{Fig:PAL} and \ref{Fig:TEE} for details 
	    about the crew and the journey). Each panel presents 
	    the same simulation achieved with a different 
	    number of iterations. More loops means a better 
	    statistical estimation of the representative result.} 
    \label{Fig:Loops}
\end{figure*}

We explored this issue and present in Fig.~\ref{Fig:Loops} four realisations of the same simulation using a different
number of iterations. The total energy expenditure per year in the vessel for a vigorously active population is 
plotted as a function of the number of iterations. In the case of a single loop, the results are noise-dominated 
and uncertainties prevail over the real median outcome. When the simulation is looped ten times the results start 
to stabilize and specific features (such as the sudden decrease in population at 750 years due to the 
catastrophic event on-board) start to appear. Looping the simulation one hundred times gives us a median value 
that is no longer subject to high statistical fluctuations, except at the beginning of the journey where the 
population is smaller than in the end, hence initial statistics are still poor. Nevertheless the results at the 
end of the interstellar trip are already significant. Finally, looping the simulation one thousand times gives 
us access to a very smooth median outcome. Statistical fluctuations are almost non-existent and we can safely 
conclude that one thousand iterations are perfectly sufficient. 

To better understand what this represents, looping this simulation a thousand times means that about one million
humans have been simulated. The total number of humans simulated to obtain Fig.~\ref{Fig:TEE_vs_Crew} corresponds 
to approximately two hundred million, three times the current population of France \cite{INSEE2018}.

\end{document}